\documentclass[oldversion]{aa}

\usepackage{txfonts}

\usepackage{graphicx}
\usepackage{natbib}
\usepackage{amsbsy}

\newcommand{\vecV}{{\vec V}}
\newcommand{\vecr}{{\vec r}}
\newcommand{\vecu}{{\vec u}}
\newcommand{\reff}{{\mathrm{eff}}}
\newcommand{\rd}{{\mathrm{d}}}
\newcommand{\vk}{{v_\mathrm{k}}}
\newcommand{\cs}{{c_\mathrm{s}}}
\newcommand{\aur}{{\mathrm{AU}}}
\bibpunct{(}{)}{;}{a}{}{,}

\begin{document}

\title{On gas drag in a circular binary system}

\author{P. Cieciel\c{a}g \inst{1, 3} \and S. Ida \inst{2} \and
A. Gawryszczak \inst{1} \and A. Burkert \inst{3} }

\offprints{P. Cieciel\c{a}g}

\institute{
Copernicus Astronomical Center, Bartycka 18, 00-716 Warsaw, Poland\\
\email{pci@camk.edu.pl}
\and
Department of Earth and Planetary Sciences, Tokyo Institute of
Technology, Ookayama, Meguro-ku, Tokyo 152-8551, Japan
\and
University Observatory Munich, Scheinerstr. 1, 81679 Munich, Germany
}

\date{}

\abstract{We investigate both analytically and numerically the motion
of massless particles orbiting primary star in a close circular binary
system with particular focus on the gas drag effects. 
These are the first calculations with particles ranging in size from 
1\,m to 10\,km, which account for the presence of a tidally perturbed
gaseous disk.
We have found numerically that  the {\em radial} mass transport by 
the tidal waves plays a crucial role in the orbital evolution of particles.
In the outer region of the gaseous disk, where its perturbation is
strongest, the migration rate of a particle for all considered sizes 
is enhanced by a factor of 3 with respect to the axisymmetric disk in
radial equilibrium. Similar enhancement is observed in the damping
rate of inclinations.
Numerical results are confirmed by more general analytical
calculations that do not assume anything about origin of the radial 
flow in the disk.
We present a simple analytical argument proving that the migration
rate of a particle in such disk is enhanced due to the enhanced mass
flux of gas colliding with the particle. 
Thus the enhancement factor does not depend on the sign of
the radial gas velocity, and the migration is always directed inward.
Within the framework of the perturbation theory we derive more general,
approximate formulae for short-term variations of the particle semi-major
axis, eccentricity and inclination in a disk out of radial equilibrium.
The basic version of the formulae applies to the axisymmetric 
disk, but we present how to account for departures from axial symmetry
by introducing {\em effective} components of the gas
velocity. Comparison with numerical results proves that our 
formulae are correct within several percent.
We have also found in numerical simulations that the tidal waves
introduce coherence in periastron longitude and eccentricity for
particles on neighbouring orbits. The degree of the coherence depends
on the particle size and on the distance from the primary star, being
most prominent for particles with 10\,m radius. The results are
important mainly in the context of planetesimal formation and, to a
lesser degree, during the early planetesimal accretion stage.
}


\maketitle


\section{Introduction}
Given the number of currently known extrasolar planetary systems
($\sim$200), it is tempting to draw some conclusions about their
formation from statistical correlations.
One of the basic correlations is the membership in a multiple stellar
system. About 15\% of the planets were found in multiple stellar
systems. Assuming that the planets are equally common around single
stars and in multiple systems, this is a clear discrepancy with the
observed frequency of the multiplicity of solar type stars in the solar
neighborhood, which is about 60\% \citep{Duquennoy91}.
A natural explanation is observational selection effects, because
multiple systems have been usually excluded from radial velocity searches
for planets.
Whether it is the only source of discrepancy will remain unknown until
more systematic searches for planets in multiple systems are carried
out. Such searches are currently underway \citep[e.g.][]{PHASES}.
It seems reasonable, however, to expect that the planet formation process
is influenced by companion stars.
Distant companions (say, with periastra farther than 100\,AU)
do not affect planetary systems too much, but several extrasolar
planets in binary systems with relatively small separation (less than
100\,AU) have also been discovered  \citep[and references
therein]{Eggenberger04}.
In particular, three planets have been discovered in binary systems
with separations of $\sim 20$\,AU  (HD41004A b, $\gamma$ Cep b, 
Gl86 b). These close companions must have modified the structure of the
protoplanetary disk (i.e. initial conditions for planet formation), as
well as the dynamical evolution of planetary orbits.

Although the number of extrasolar planets in multiple systems is
currently very small, some correlations seem to be statistically
significant. \citet{Zucker02} pointed out that all the most massive 
(more than 2 Jupiter masses), short-period (periods shorter than 40\,d)
planets orbit stars from binary systems. In consequence, for
planets in binaries there is no correlation between mass and period as
is observed for planets around single stars. 
\citet{Eggenberger04} showed that orbital eccentricities
tend to be very low for short-period planets.
All those differences indicate that the companion star is affecting
the planet formation process.

To address planet formation in binary systems, studies of the orbital 
evolution of planetesimals have been done 
\citep[e.g.][]{Hep78, Whitmire98, Marzari00, Quintana02, Thebault04,
Thebault06}. In the classical scenario without a gaseous disk, the
secular perturbations from the binary companion pump up orbital
eccentricities of planetesimals and decelerate their accretion by
reducing the gravitational focusing factor.
Also, if the relative velocities exceed the escape velocity from
their surface, collisions result in disruption rather than
coagulation \citep[e.g.][]{Agnor04}, so that planetesimal
accretion is inhibited. 
\citet{Marzari00} included the effects of uniform gas drag in an
eccentric binary system and found strong periastron alignment of 
equal-size planetesimals.
If the periastra are aligned, the relative velocities are kept low in
spite of high eccentricities, and planetesimal accretion is not
inhibited. However, \citet{Thebault06} pointed out that
the alignment angle depends on planetesimal size, and if the size
distribution is introduced, the relative velocities between 
particles of different sizes are typically prohibitive for the
collisional growth. 
An important result of these works is that even a small gas-drag
force can significantly change the growth rate of planetesimals if
combined with perturbations of the companion star.
The weak point of these studies is the assumption that the gaseous
disk is not perturbed, but remains stationary and axisymmetric.
It is known \citep[e.g.][]{PapPringle77, LinPap79, GT79}, however,
that the companion induces tidal waves in the gaseous disk,
which can evolve into strong spiral shocks. The density and velocity
of the gas are then strongly perturbed, and the drag force acting on
solid particles is different than in the stationary, axisymmetric case.

To date no calculations of particle motion in a disk with tidally
induced spiral waves have been done. 
In a series of papers we will explore this subject, both analytically
and numerically studying the orbital evolution and accretion of 
particles in disks perturbed by a companion star.

In this paper we consider the case of non-interacting particles
orbiting in a circumprimary disk, with the perturbing companion star on
a {\em circular} orbit. The particles range in size from 1\,m to
10\,km, so our results apply to the planetesimal formation and early
accretion phases.
At first glance, the circular case may not seem interesting because
the gravity of the companion does not induce secular effects on
the particle orbit, like eccentricity forcing or 
periastra libration. However, we show that effects similar to those
from an eccentric binary are also observed in the circular case if the
perturbations of the gaseous disk are included. Furthermore, we
show that the orbital evolution of particles in such a system is
significantly different than in the unperturbed, axisymmetric disk.
The circular case is a very good starting point since it allows us to
understand the sole effect of spiral waves in the gaseous disk. 
The eccentric binary case will be discussed in next paper. 

The paper is organized as follows. 
In Sect. \ref{sec:comp_method} we present computational methods 
and input physics.
Some important properties of the gaseous disk are discussed in Sect.
\ref{sec:gas_disk}.
In Sect. \ref{sec:orb_evol} we investigate analytically and
numerically the orbital evolution of a single particle. 
Section \ref{sec:coherence} is devoted to relative shapes and
alignment of neighbouring orbits of particles.
In Sect. \ref{sec:summary} we summarize and discuss the results.


\section{Computational method \label{sec:comp_method}}

The problem we investigated involves the solution of both gas and particle 
equations of motion. One approach would be to combine hydro and N-body 
schemes into a single numerical code, however, it was enough for our purpose
to perform two-stage simulations. We exploited the fact that in the circular 
binary system the pattern of spiral waves in the gaseous disk is
quasi-stationary in the frame co-rotating with the secondary star. By
quasi-stationarity we mean here that the time scale of its evolution is
longer than the time scale of the evolution of N-body particles.
In the first step we obtained such a quasi-stationary model of the gaseous disk 
and then we fed it to the N-body code. In this way we eliminated the temporal 
evolution of the gas. This was very desirable since our goal was to 
investigate generic effects of the spiral shocks on the motion of the 
solid bodies and not to perform realistic simulations of the 
particle growth in the binary system, a task we leave for future work.

The binary system we simulated consists of primary and secondary stars
of equal mass, $M_p=M_s=1 M_{\odot}$ on a fixed circular orbit with
the semi-major axis $a=23.4$\,AU. The implied orbital period is close
to 80\,yr. 
The gaseous disk and particles are orbiting a primary
star. We chose these parameters because, apart
from the eccentricity, they are close to the $\alpha\ Centauri$ system
investigated in \citet{Marzari00}, and it will enable us to compare
their results with ours, especially in the follow-up paper about the
eccentric binary case.
Note that the self-gravity is not included in either fluid or particle
simulations , so it is possible to scale the models to any size.

\subsection{Hydrodynamical simulation}
Our model of the gaseous disk was evolved using an adaptive mesh refinement 
(AMR) code, FLASH \citep{FLASH00}. As a hydro solver we employed direct 
Eulerian PPM \citep{CW1984} scheme modified to conserve angular momentum. 
The PPM scheme combines high-order spatial interpolation with a Riemann solver 
and shock-capturing method that results in low numerical viscosity and
sharp shock profiles. This makes PPM particularly useful for all applications
requiring accurate transport of momentum in a supersonic flow environment.

The code solved Euler equations in 2D polar coordinates with the origin located 
at the primary star:
\begin{equation} \label{eq:hydro_cont}
\frac{\partial\Sigma}{\partial t}+\nabla\cdot(\Sigma\vecV)=0
\end{equation}
\begin{equation} \label{eq:hydro_motion}
\frac{\partial(\Sigma\vecV)}{\partial t} +
\nabla\cdot(\Sigma\vecV\otimes\vecV)+\nabla P = -G\Sigma \left(
M_p\frac{\vecr}{r^3} + M_s \frac{\vecr-\vecr_s}{|{\vecr-\vecr_s}|^3} + M_s \frac{\vecr_s}{r_s^3}\right)
\end{equation}
where $\Sigma, P$, and $\vec{V}=(V_r, V_\phi)$ denote surface density,
surface pressure, and gas velocity at position $\vec{r}$. The
secondary star was located at $\vec{r}_s$ such that $r_s=a$.
The equation of state was locally isothermal (temperature was a fixed 
function of distance from the primary star), so there was no need to
solve the energy equation, and a faster isothermal version of the
Riemann solver was used.

The final model of the gaseous disk was obtained as follows: 
(1) The initial disk was set up as in the case of a single star, and
it was truncated exponentially beyond a radius slightly larger than
the expected tidal truncation radius.
(2) During the first two orbital periods of the binary system the grid
resolution was increased up to 2048x768 in $r$ and $\phi$
respectively (we used the AMR option here, but to avoid any artifacts
at the edges of the refined blocks the whole grid was always refined).
(3) The disk was evolved for an additional orbital period of the binary.

The Courant number was 0.5.
The grid extended radially from 0.4\,AU to 9\,AU -- far enough to avoid
any influence from boundaries on the most interesting region of the
outer disk where the spiral waves are strongest. We took special care
to minimize reflections at the inner and outer grid boundaries. For
this purpose we tuned the standard outflow boundary conditions in
order to reduce any discontinuities in radial direction. We also
introduced a low-density hole in the first 5 radial cells at the inner
boundary, which served as an additional buffer to damp the propagating
waves. We stress that all these measures are necessary in order to
recover the proper mass transport in the non-viscid disk.

\subsection{Orbital integration}

Our orbital integration code is based on Nbody4 \citep{Aarseth79}.
It implements the direct summation method for self-gravity, 4th-order 
Hermite scheme, and block time step \citep{Makino91a}. 
In this paper we consider motion of particles only under gravitational forces
of the two stars and gas drag force, and mutual gravitational forces
of planetesimals are neglected.
Inter-particle forces and collisional accretion will be included in
future papers.

In the reference frame located at the primary star, the corresponding
equation of motion for particle $i$ with mass $m_i$ reads:
\begin{equation} \label{eqmot}
\ddot{\vecr_i} = -G (M_p+m_i)\frac{\vecr_i}{r_i^3} - G M_s
\frac{\vecr_i-\vecr_s}{|\vecr_i-\vecr_s|^3} - G M_s
\frac{\vecr_s}{r_s^3} + {\vec{f}}_{drag,i},
\end{equation}
where $\vecr_s$ denotes the secondary star position. Components on the
right hand side of Eq.~\ref{eqmot} represent the gravity of the primary and  
secondary, the indirect term accounting for acceleration of the 
primary relative to the center of mass and gas drag force per unit
mass, respectively. For the latter we adopt a simple formula:
\begin{equation} \label{eq:drag_force}
\vec{f}_{drag,i} = -A \rho |\vecu|\vecu , ~~~~~~~A=\frac{1}{2m_i}C_\mathrm{D} \pi s_i^2,
\end{equation}
where $s_i$ is the particle radius, $\rho$ -- gas density, 
$C_\mathrm{D}$ -- drag coefficient, and $\vecu$ is the particle
velocity relative to the gas. The factor $A$ is constant in our simulations.
Denoting the particle's velocity with $\vec v$ we have
\begin{equation} \label{eq:urel}
u_r = v_r-V_r ~,\quad\quad u_{\phi}=v_{\phi}-V_{\phi}.
\end{equation}
We used values of $C_\mathrm{D}=1.4$ and internal density of particles
$\rho_p=2\, \mathrm{g/cm}^3$. The Hermite scheme also requires the value of
$\dot{\vec{f}}_{drag}$, which is a minor correction that was accounted for
using numerically calculated values of $\dot{\rho}$ and $\dot{\vecu}$.

We took special care to include gas-drag effects accurately yet efficiently in 
calculations. 
At the beginning of the orbital calculation, the grid data containing
gas density and velocity were read in. During the simulation 
the data were rotated to match the current position angle of the secondary 
star (spiral pattern of the gas co-rotates with the companion), and the
bi-linear interpolation was used to find the gas state at an arbitrary position
of the particle.

The time step was variable but limited to a maximum of $1/(2\pi\cdot64)$\,yr.
We tested the code with basic problems like conservation of the Jacobi
constant, and properly recovered more complex results like runaway growth 
\citep{KokuboIda96} and gas drag in a uniform disk \citep{Inaba01}.

\subsection{The models}
\label{ch:Models}
Each model here is composed of the two components: gaseous disk and
stellar system configuration. We used the following three gaseous
disk configurations:
\begin{itemize} 
\item Axisymmetric disk in radial equilibrium or in a short equilibrium
axisymmetric disk. Characterized by a simple power-law density and
temperature radial profiles, the radial gas velocity is zero.
\item Axisymmetric disk in radial non-equilibrium or in a short
non-equilibrium axisymmetric disk. It is similar to the equilibrium
axisymmetric disk, but the radial gas velocity is artificially
set to a non-zero value.
\item Non-axisymmetric disk, resulting from the evolution of an
initially equilibrium axisymmetric disk in a circular binary
system. It develops a spiral wave pattern and is naturally in radial
non-equilibrium.
\end{itemize}
The stellar configuration can simply be either a {\bf single star} or a {\bf
circular binary system}. 
Sometimes we refer to the models using abbreviations presented in the
Table \ref{tab:models}.
\begin{table}[!h]
\caption[]{Abbreviations of presented models.}
\label{tab:models} 
\begin{tabular}{|l|c|c|}
\hline
~&single star&binary system\\
\hline
equilibrium axisymmetric disk&EA1&EA2\\
non-equilibrium axisymmetric disk&NA1&NA2\\
non-axisymmetric disk&W1&W2\\
\hline
\end{tabular}
\end{table}

We note here that only models EA1 and W2 are physically
consistent, while only model EA2 has been investigated by other authors.
Thus we concentrate on differences between models EA2 and W2.
The other models are used only as a support in understanding the observed
effects.

\section{Gaseous disk \label{sec:gas_disk}}

\subsection{Disk parameters and structure}
Our numerical method requires the gaseous disk to be in a state close to
stationarity. It should also be minimally biased by numerical effects
and should adequately recover all deviations from the Keplerian flow.
To that end we used a fairly simple isothermal model, which is 
nonetheless is close to the minimum mass solar nebula \citep{Hayashi81}.

The equation of state,
\begin{equation}
P=\Sigma \cs^2,
\end{equation}
is locally isothermal, with the local sound speed $\cs$, given by
the vertical hydrostatic equilibrium condition
\begin{equation}
\cs = \frac{h}{r} v_{k},
\end{equation}
where $v_{k}=\sqrt{GM_p/r}$ is the Keplerian velocity and $h$ the
local half-thickness of the disk. \citet{Blondin00} has shown that, in
sufficiently cold disks, the spiral waves at the outer edge of the disk
may become unsteady. We found experimentally that for the following
height profile
\begin{equation}
h_r \equiv \frac{h}{r} = 0.05\cdot \left(\frac{r}{1\,\aur}\right)^{0.5}
\end{equation}
the spiral pattern stays stable everywhere.
Note that, although our equation of state is locally isothermal, such
$h_r$ results in global isothermality with $\cs=0.05\sqrt{GM_p/(1\,\aur)}$. 
The corresponding Mach number ($\vk/\cs$) is 31.4 at the inner disk
edge ($r$=0.9\,AU), which falls to 7.6 at the outer edge ($r$=9\,AU).

The initial profile of the surface density was given by the power law
\begin{equation}
\Sigma_i = \Sigma_0 \left(\frac{r}{1\,\aur}\right)^{-1.5},
\end{equation}
which is close to the MMSN model. Since self-gravity is not included in
hydro simulation, the normalization $\Sigma_0$ can be arbitrary.
To calculate the drag force (Eq.~\ref{eq:drag_force}) 
during orbital integration, the evolved surface density $\Sigma(\vec{r})$ was converted 
to the volume density and normalized as follows:
\begin{equation}
\label{eq:voldens}
\rho = 2\cdot 10^{-9} ~\frac{\Sigma/\Sigma_0}{h(r)/1\,\aur} = 
       2\cdot 10^{-9} ~\frac{\Sigma}{\Sigma_0}~ \left(\frac{r}{1\,\aur}\right)^{-1.5} \mathrm{[g/cm^3]}.
\end{equation}
The initial angular velocity was set to the equilibrium one for the given pressure gradient
\begin{equation}
\label{eq:vang2d}
V_\phi = \vk + \frac{1}{2} \frac{\partial \textrm{ln}\,P}{\partial \textrm{ln}\,r} \frac{\cs^2}{\vk},
\end{equation}
and the radial velocity was set to 0.

The initial conditions described above represent our equilibrium
axisymmetric disk configuration.
This configuration was evolved numerically for three binary orbital periods,
which is enough to develop a quasi-stable spiral pattern.
This evolved disk will be referred to as the non-axisymmetric disk.

Here we have to point out that the 2D approximation
introduces a certain inconsistency into both models of the gaseous disk.
The problem with 2D hydrodynamical simulations is that the velocity
field cannot be directly linked to the 3D one. The 2D velocity given by
Eq. \ref{eq:vang2d} assures radial equilibrium 
for a given gradient of the {\it vertically averaged  pressure}.
Note, however, that it is neither the vertically averaged velocity
nor velocity in the equatorial plane. Proper, equatorial plane velocity is
given by
\begin{equation}
\label{eq:vang3d}
V_\phi = \vk + \frac{1}{2} \frac{\partial \textrm{ln}\,p}{\partial \textrm{ln}\,r} \frac{\cs^2}{\vk}
\end{equation}
where $p=\rho \cs^2$ denotes pressure in the equatorial plane. Both
formulae give different results since, in general, the gradient of $p$ is
different from the gradient of $P$. 
Unfortunately, the simulated 2D velocity cannot be transformed 
to the equatorial plane velocity. Thus an inconsistency
arises: we use 2D velocities, while the density is converted to 3D
one (Eq. \ref{eq:voldens}). We decided that it is better to use
Eq. \ref{eq:vang2d} for the axisymmetric disk model although in principle
Eq. \ref{eq:vang3d} should be used: our results may not be accurate
quantitatively, but at least we can compare both models qualitatively.
Furthermore we expect {\it relative} results from both models to be less
affected than absolute ones. 

For the later considerations it is useful to define two dimensionless velocities
which describe deviations from the Keplerian flow:
\begin{equation}
\eta = (\vk-V_{\phi})/\vk,
\end{equation}
\begin{equation}
\kappa = -V_r/\vk.
\end{equation}
In other words, these are velocity components of a large particle moving 
on a circular, Keplerian orbit relative to the gas.

\begin{figure}
\resizebox{\hsize}{!}{
\includegraphics[]{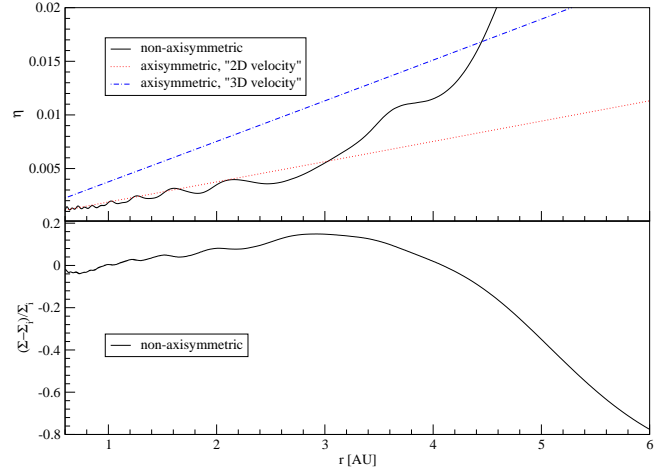}
}
\caption{ \label{fig:rad_prof} Top panel: radial profiles of
angle-averaged $\eta$ in non-axisymmetric disk (solid), axisymmetric
disk with ``2D velocity'' (Eq. \ref{eq:vang2d}, dotted), and axisymmetric
with ``3D velocity'' (Eq. \ref{eq:vang3d}, dot-dashed).
Bottom panel: radial profile of angle-averaged surface overdensity,
$(\Sigma-\Sigma_i)/\Sigma_i$, in a non-axisymmetric disk.}
\end{figure}
To reveal the influence of spiral waves on the particle motion,
we have to compare results obtained in the axisymmetric disk with
those obtained in the perturbed disk at a radius where angle-averaged
values of $\rho, V_r$, and $V_\phi$ are comparable.
The upper panel of Fig. \ref{fig:rad_prof} shows the radial profiles of $\eta$ 
averaged over the full angle for three disk models: non-axisymmetric,
axisymmetric employing formula \ref{eq:vang2d}, and 
axisymmetric employing formula \ref{eq:vang3d}.
As we see, the profiles of the two first models are comparable up to
a distance of 3\,AU from the primary star. Outside of this region the
deviations from the Keplerian flow grow substantially. Since the
spiral waves are strongest in the outer region of the disk, in the next
section we will trace the motion of the particle placed initially at
3\,AU. The dash-dotted curve illustrates why the ``3D velocity'' is
not suitable for our comparison. The density in the non-axisymmetric
model at 3\,AU has grown during the simulation time by roughly 15\% with
respect to the initial model (see lower panel of Fig.
\ref{fig:rad_prof}). Since the migration speed due to gas drag scales
linearly with the density, this difference can be easily accounted for
in comparisons with the axisymmetric disk.

\subsection{Radial transport of gas}
\begin{figure}
\resizebox{\hsize}{!}{
\includegraphics[]{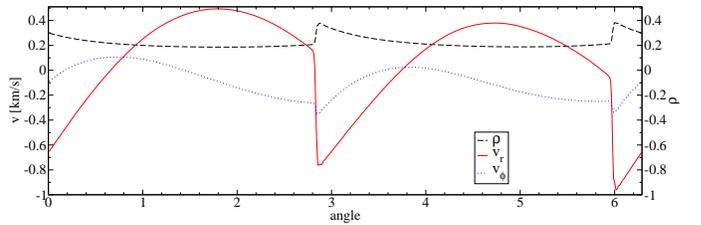}
}
\caption{\label{fig:ang_prof} Angular cross-sections of velocity
components and density at $r$=3\,AU.}
\end{figure}

When the spiral density waves are excited in the disk, it is no longer in
radial equilibrium. Because the
spiral pattern is rotating slower than the local Keplerian velocity, the
dissipation at the shock leads to a decrease in angular momentum of the
orbiting gas and its radial mass transport. Figure \ref{fig:ang_prof}
shows the angular cross sections of velocity components and density at
3\,AU in the non-axisymmetric disk. 
Indeed, the angle-integrated mass
flux calculated from those profiles is negative (inward). It can be
interpreted as the result of an effective viscosity that we define as
the turbulent viscosity necessary to cause the same radial mass flux.
In the standard $\alpha$-disk theory, the kinematic viscosity $\nu$ is
expressed in terms of a dimensionless parameter $\alpha$ as
$\nu=2/3\alpha \cs h$. Assuming a steady accretion disk,
i.e. $\dot{m}=-3\pi \nu \Sigma$ independent of $r$, we can
parametrize the mass accretion rate with the effective value of
$\alpha_\reff$:
\begin{equation}
\dot{m}=2\pi r V_r \Sigma = -2 \pi \alpha_\reff \cs h \Sigma.
\end{equation}
For the non-axisymmetric disk, the above equations must be averaged
over the azimuthal angle, and finally the effective $\alpha$-viscosity is
defined as
\begin{equation}
\alpha_\reff = - \frac{2 \vk V_\mathrm{r, eff}}{3 \cs^2},
\end{equation}
where
\begin{equation}
V_\mathrm{r, eff} = \frac{\int_0^{2\pi}\Sigma V_\mathrm{r} \,
\mathrm{d}\phi}{\int_0^{2\pi}\Sigma \, \mathrm{d}\phi}.
\end{equation}

It has been shown analytically \citep{Spruit87} and numerically
\citep{Blondin00, Rozyczka93} that the mass transport by tidal waves can be very
effective. For reasonable disk parameters in close binary systems,
the effective $\alpha$-viscosity in the outer parts of the disk can
easily reach 0.1 or even more. It is hard to produce such high values
by ordinary turbulent viscosity.

\begin{figure}
\resizebox{\hsize}{!}{
\includegraphics[]{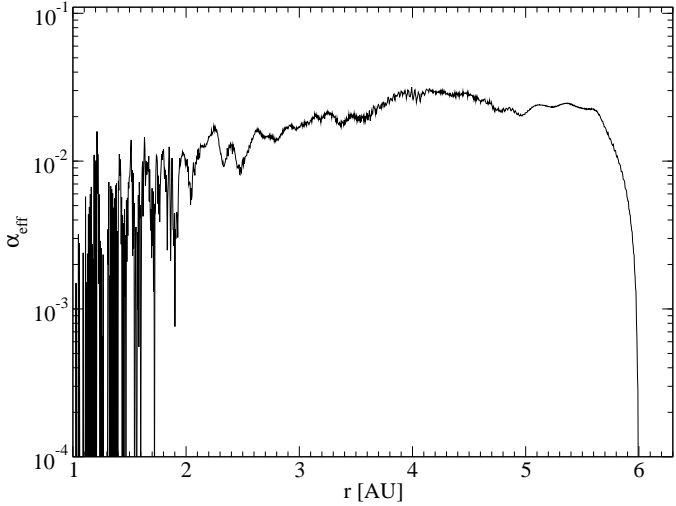}
}
\caption{\label{alpha} Radial profile of the effective $\alpha$-viscosity.}
\end{figure}
The radial profile of $\alpha_\reff$ in our non-axisymmetric disk is shown
in Fig. \ref{alpha}. Close to the inner edge of the disk, it
oscillates strongly because the spiral waves are very tightly wound,
and the radial grid resolution is insufficient for resolving
them. However in the region close to 3\,AU, the resolution
is sufficient, and the effective $\alpha$ may be easily found.

\section{Evolution of orbital elements \label{sec:orb_evol}}

\subsection{Non-equilibrium axisymmetric disk: analytical calculations}

The rate of change of orbital elements of a particle experiencing
gas drag was calculated by \citet[][hereafter A76]{Adachi76}. 
Their results cannot be applied, however, applied to a disk in the
binary system because the authors assumed that the disk is
axisymmetric and stays in radial equilibrium ($V_r=0$). In the binary
system the disk is neither axisymmetric nor in radial equilibrium, so
the evolution of orbital elements may be very different. In this
subsection, we calculate analytically the evolution of orbital elements
of a particle moving through the gaseous disk, which is not in radial
equilibrium, while still neglecting the action of the companion on
particles. In other words, we extend the A76 work to the case of
non-zero radial gas velocity.

First, we consider the simplest case of the particle on a nearly
circular, non-inclined orbit.
Let $u=\sqrt{u_r^2+u_\phi^2}$ be the value of the total relative velocity
between particle and gas, where ($u_r$, $u_\phi$) are radial and 
angular components, respectively.
The particle loses specific angular momentum only due to angular
component of the drag force, and for small drag (when the orbit stays 
nearly circular) its loss rate can be approximated as
\begin{equation}
\frac{\rd(\vk a)}{\rd t} = \frac{1}{2}\vk \frac{\rd a}{\rd t} \approx 
-\frac{u_\phi \cdot a}{\tau}
\end{equation}
where $\vk$ is the Keplerian velocity at radius $a$, and $\tau$ is 
the stopping timescale:
\begin{equation}
\tau = \frac{u_\phi}{A\rho u_\phi u}.
\end{equation}
Thus the orbit decay rate is given by the approximate formula
\begin{equation}
\label{eq:dadt_approx}
\frac{\rd a}{\rd t}\approx -2A\rho a \frac{u_\phi}{\vk}u.
\end{equation}
There are two important factors here: $u_\phi$ and $\rho u$. The particle
loses angular momentum when colliding with the gas at relative
velocity $u_\phi$, but the mass flux of this gas is $\rho u$, which is
how the radial gas velocity enhances the migration rate.
For large enough particles moving with Keplerian velocity, we can
further write
\begin{equation}
\label{eq:dadt_eta}
\frac{\rd a}{\rd t} \approx -2A\rho a\eta u.
\end{equation}

In appendix A we calculate the evolution of orbital elements for the general
case of eccentric and inclined orbits within the framework of
perturbation theory. In the limit of a circular, non-inclined orbit, the
general formula for orbitally averaged $da/dt$ (Eq. \ref{eq:dadt})
reduces exactly to Eq. \ref{eq:dadt_eta}.

\begin{figure}
\resizebox{\hsize}{!}{
\includegraphics[]{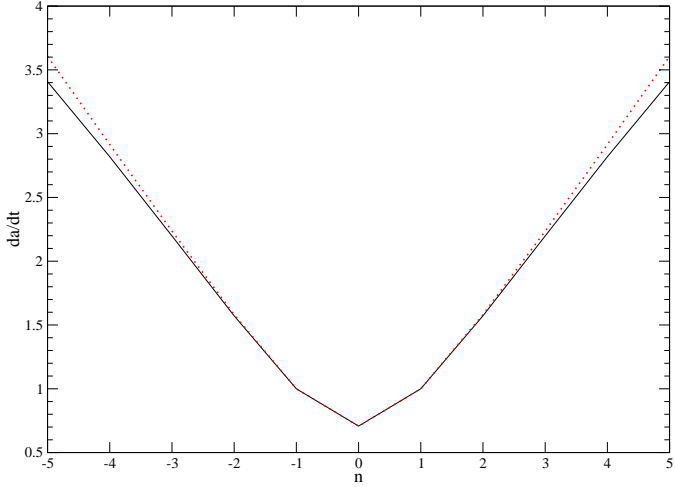}
}
\caption{\label{fig:adot-n} Dependence of the particle migration rate
(in units of $\dot{a}(n=1)$) on the radial gas velocity. Solid line -
simulation; dotted line -- analytical approximation (Eq. \ref{eq:adot-n}).}
\end{figure}

We would like to turn the reader's attention to the dependence on the 
{\em value} of total relative velocity in formula (\ref{eq:dadt_approx})
(for details, see Appendix).
In particular, in the limit of high radial velocity, $u_r \gg u_\phi$, 
we have a very surprising relation:
\begin{equation}
\frac{\rd a}{\rd t} \propto -\vert u_r \vert.
\end{equation}
The migration rate is proportional to the absolute value of gas radial
velocity; i.e., the particle migrates inward regardless of the direction
of the radial gas flow! In order to test this result numerically
we measured particle migration rates in the axisymmetric disk, while
artificially varying the radial gas velocity.
The particle was initially on a circular orbit with components of 
relative velocity $u_\phi$ and $u_r = V_r = n \cdot u_\phi$, where 
$n$ was an integer number between -5 and 5.
Figure \ref{fig:adot-n} shows measured and predicted migration rates 
(normalized to $\dot{a}(n=1)=1$) as a function of $n$. The predicted
curve is given by the formula
\begin{equation}
\label{eq:adot-n}
\dot{a}(n)=\sqrt{\frac{1+n^2}{2}}.
\end{equation}
Indeed, the migration rates measured in the simulation for $n$ and $-n$
are the same within 1\%. Also the agreement with prediction 
(Eq. \ref{eq:adot-n})
is very good, especially for small $n$. For higher values of $|n|$, formula 
($\ref{eq:adot-n}$) deviates slightly from results of the simulation
because the orbit differs more and more from the assumed circular
shape.
In fact, changes in the semi-major axis and eccentricity are coupled
so must be considered together. Even though the radial gas velocity
does not change the angular momentum of the particle directly, it does
change its eccentricity, which in turn affects the decay rate of the
semimajor axis  (see Eqs. \ref{eq:dadt_avg}-\ref{eq:deidt_avg}).

The above result concerns an idealized axisymmetric disk and aims to
enhance basic effects of radial gas flow (regardless of its source),
as well as to provide a proof of code validity. 

\subsection{Non-axisymmetric disk: numerical and analytical results}

In this section we present the results of the orbital calculations in
model W2 and compare them with models EA1, EA2 and W1. 
We also derive an analytical approximation to the perturbative formulae
\ref{eq:dadt}-\ref{eq:deidt} for the case of non-axisymmetric disk and
compare it with numerical results from model W2.

\begin{figure}
\resizebox{\hsize}{!}{
\includegraphics[]{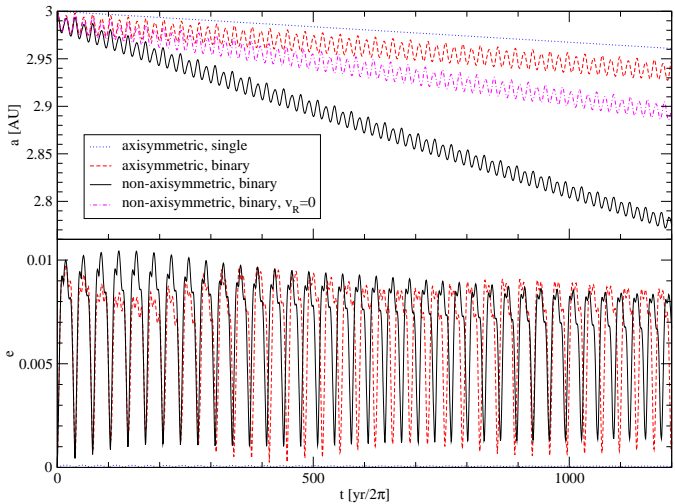}
}
\caption{\label{fig:ae-t} Semimajor axis (upper panel) and eccentricity
(lower panel) versus time for a 10\,m particle in models: W2 (solid),
EA2 (dashed), EA1 (dotted) and W2 with radial gas velocity
set to zero (dash-dotted).}
\end{figure}
As a first example, we consider particle of radius 10\,m on
an initially non-inclined, circular orbit of radius 3\,AU. 
We checked empirically that in all our models the particle of this
size migrates by less than 0.2\,AU within the simulation time, so that
the parameters of the gaseous disk can be regarded as roughly
constant. This enables us to compare the migration speed between
different disk models.
The upper panel of Fig. \ref{fig:ae-t} shows the evolution of
the particle's semimajor axis in models W2, EA2, and EA1.
As we see, the difference in migration speed of a particle in the
non-axisymmetric disk is substantially enhanced with respect to the
other two cases, roughly by a factor of three.
The lower panel of Fig. \ref{fig:ae-t} shows the corresponding evolution 
of eccentricity. Oscillations with the synodic period of the companion
star are clearly seen. We checked that the oscillations are not damped 
by gas drag for particles larger than 10\,m. The amplitude of oscillations
depends on the semimajor axis, and the mean eccentricity is roughly
the same in both models with a companion star (only a very small
decrease is observed with decreasing $a$). For this reason the
eccentricity cannot be responsible for the difference in the migration speed.
We have already shown analytically that this relatively fast migration 
can be induced by the radial gas velocity.
In order to test this prediction numerically, we performed an orbital
calculation with the non-axisymmetric disk in which the radial gas
velocity was set artificially to zero. The corresponding curve in Fig.
\ref{fig:ae-t} clearly proves that the radial gas flow is the main
factor responsible for the accelerated particle migration. Another two
factors are responsible for the remaining part of the difference from
the axisymmetric model. First, in the non-axisymmetric case the
effective value of $\eta^2$ is higher due to weighting by the density
(see discussion in the next paragraph). Second, due to the
dissipation in spiral waves, the mean density in the simulation has
increased by 15\% with respect to the axisymmetric model (see
Fig. \ref{fig:rad_prof}).

How well do the above numerical results agree with the analytical
approximations? The comparison is straightforward for the equilibrium
axisymmetric disk. In that case formula \ref{eq:dadt}
reduces to the original formula 4.21 from A76. 
We have $a/\tau_0=0.51$ [2$\pi$\,AU/yr] in our disk model
for the 10\,m particle on the orbit with $a=3$\,AU. 
Furthermore, $\eta=0.0056$, and the measured mean eccentricity is
$e=0.006$. For those parameters the analytically predicted migration
velocity is within 1\% of the value measured in the simulation:
$4.3\cdot10^{-5}$ [2$\pi$\,AU/yr]. This proves the accuracy of our 
orbital integration.

Application of formula $\ref{eq:dadt}$ to the non-axisymmetric disk is
not as straightforward. We have found that simply inserting angular
averages of $\eta$ and $\kappa$ leads to quite a large discrepancy with
the value measured in simulation. This is because the density in our
non-axisymmetric disk {\em is} correlated with the velocity so the
approximation \ref{eq:uF_approx} is not justified. Here we derive a
new version of the formulae \ref{eq:dadt}-\ref{eq:deidt} that takes
such a correlation into account. To that end we modify approximation 
\ref{eq:uF_approx} by also detaching $\rho$ from $F$ and averaging 
its product with $u$ separately:
\begin{equation}
\label{eq:uF_approx2}
\langle \rho uF \rangle = \{\langle (\rho u)^2 \rangle\}^{1/2} \langle F
\rangle,
\end{equation}
where the orbital average, $\langle \rangle$ is defined by Eq. \ref{eq:orbavg}.
This leads to the following formulae:
\begin{equation}
\label{eq:dadt_avg}
\frac{\tau_0'}{a}\left\langle\frac{\rd a}{\rd t}\right\rangle = 
-2\left[ \left(\frac{5}{8}-\kappa_\reff^2\right) e^2 + 
\frac{1}{2}i^2+ \eta_\reff^2 + \kappa_\reff^2 \right] ^{1/2}\eta
\end{equation}
\begin{equation}
\label{eq:deidt_avg}
\frac{\tau_0'}{e}\left\langle\frac{\rd e}{\rd t}\right\rangle = 
2\frac{\tau_0'}{i}\left\langle\frac{\rd i}{\rd t}\right\rangle = 
\left[ \left(\frac{5}{8}-\kappa_\reff^2\right) e^2 + 
\frac{1}{2}i^2+ \eta_\reff^2 + \kappa_\reff^2 \right]^{1/2},
\end{equation}
which are very similar to Eqs. \ref{eq:dadt}-\ref{eq:deidt}; but
variables $\eta$ and $\kappa$, which enter the formula for the total relative
velocity (Eq. \ref{eq:usq}), have been changed to the {\em
effective} values
\begin{equation}
\label{eq:etaeff}
\eta^2 \rightarrow \eta_\reff^2 = \frac{\overline{(\rho \eta)^2}}{\overline{\rho}^2}
\end{equation}
\begin{equation}
\label{eq:kappaeff}
\kappa^2 \rightarrow \kappa_\reff^2 = \frac{\overline{(\rho \kappa)^2 }}{\overline{\rho}^2}.
\end{equation}
The bar denotes here averaging over the full angle at radius $r=a$ in
contrast to the orbital average defined by Eq. \ref{eq:orbavg}. 
Also the characteristic time scale now depends on the averaged
density:
\begin{equation}
\tau_0'=\frac{1}{A\, \overline{\rho}\, \vk(a)}.
\end{equation}
We note that $e$ and $i$ are not translated to the effective values
because we have assumed that they remain constant during one orbital
period. This might not be fulfilled for small particles that are more
strongly coupled to the gas. Fortunately, the effect becomes noticeable
only at the lower limit of the size range for which the drag law we used
is applicable. In analogy to Eq. \ref{eq:kappaeff}, we can define
the effective $\alpha$-viscosity, and its relation to $\kappa_\reff$ is
\begin{equation}
\kappa_\reff=\frac{3}{2}h_r^2 \sqrt{\overline{ \alpha_\reff^2 }}.
\end{equation}

Now we are in a position to test this result with numerical simulations.
From the model W2 we measured mean $e=0.006$,  $\eta=0.0056$, 
and effective $\kappa_\reff=0.027$, $\eta_\reff=0.01$.
For these parameters, the predicted migration velocity is $1.9\cdot 10^{-4}$
[2$\pi$\,AU/yr] in comparison to $1.8\cdot 10^{-4}$ [2$\pi$\,AU/yr]
measured from the simulation. Taking the number of
approximations that had to be done in the analytical derivation into
account, this agreement is indeed excellent.

\begin{figure}
\resizebox{\hsize}{!}{
\includegraphics[]{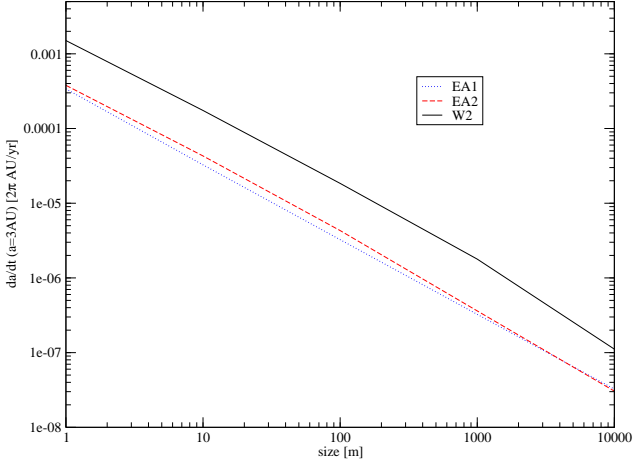}
}
\caption{\label{fig:dadt-s} Migration rate as a function of the particle
size for models: W2 -- solid line, EA2 -- dashed line, and EA1 --
dotted line.}
\end{figure}
In Fig. \ref{fig:dadt-s} we display the migration rates measured in
simulations as a function of the particle size. As we see, the
presence of the companion star does not change the migration rate in
axisymmetric disks, because the excited eccentricity oscillations are 
small in comparison to the $\eta$ parameter that plays the major role.
In the non-axisymmetric disk parameter $\kappa_\reff$ clearly
dominates other factors ($e, i, \eta_\reff$), and thus for all
considered sizes migration speed is enhanced by roughly the same
factor of 3. Of course this factor will change as $\kappa_\reff$ is
changing. The value of $\kappa_\reff$ depends on the position in the
disk, disk properties, and binary separation. In particular, for a given
disk model, it decreases with decreasing $r$, because the waves are
weaker in the inner disk. For a discussion of the dependence of
$\alpha_\reff$ on the disk density profile and temperature (for isothermal
models), please refer to \citet{Blondin00}.

\begin{figure}
\resizebox{\hsize}{!}{
\includegraphics[]{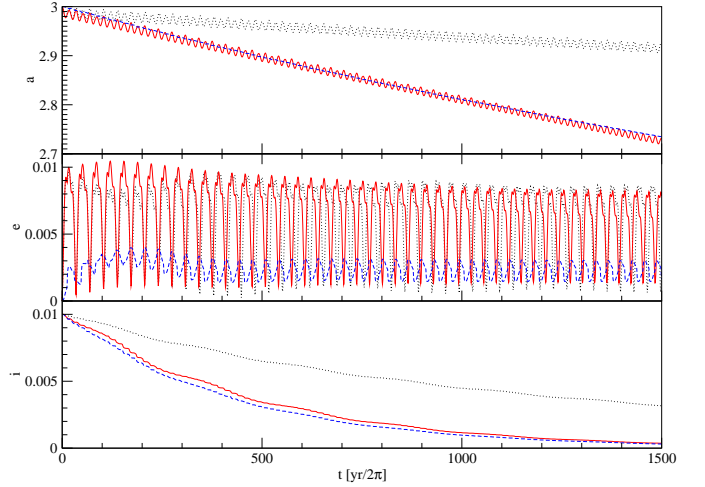}
}
\caption{\label{fig:aei-t} Evolution of orbital elements for a
particle on an initially circular, inclined orbit in models: W2 (solid
line), EA2 (dotted line), and W1 (dashed line).}
\end{figure}

To test the formula for inclination damping
(Eq. \ref{eq:deidt_avg}), we placed a particle on an initially circular
orbit at 3\,AU with inclination 0.01.
The evolution of its orbital elements in models W1, W2, and EA2 is
shown in Fig. \ref{fig:aei-t}. Clearly, the inclination decreases
faster in the non-axisymmetric disk. The initial difference in damping
rate, measured over interval $\delta t=50\,yr/2\pi$, is about three
times higher than in model EA2. In order to make a comparison with the
analytical approximation (which does not account for the presence of a
secondary), we ran one more model with the non-axisymmetric disk but
with the gravity of the secondary star switched off (W1; see
Fig. \ref{fig:ae-t}).
The difference between models W1 and W2 is
very small, which is expected since the companion star influences the
inclination rather weakly. We measured the initial slope in
model W1 to be $2.2\cdot 10^{-5}\,[2\pi/yr]$, while the approximate
Eq. \ref{eq:deidt_avg} gives $2.5\cdot 10^{-5}\,[2\pi/yr]$, which
makes a 12\% difference. Althrough worse than for the semimajor axis,
this accuracy is still very good for an approximate formula. 

The eccentricity of the particle in a binary system is set by the balance
between forcing from the secondary and drag damping. As it was
already shown in Figure \ref{fig:ae-t}, the eccentricity of a particle
on an initially circular orbit is immediately forced to oscillate with
a roughly constant amplitude (0.01 at 3\,AU). 
Comparison of the analytical formula for damping of $e$
(Eq. \ref{eq:deidt_avg})  with simulation results in model W2 would
require setting up the particle with initial eccentricity
substantially higher than 0.01. Since it is rather unlikely for a
small particle to have such eccentricities, and in addition analytical
approximation may not work well in this regime, we will not present it.
We note only that the comparison of eccentricity between models W1
and W2 in Fig. \ref{fig:aei-t} reveals that the spiral waves are only
responsible for the lower limit of eccentricity in model W2, while the
remaining contribution comes from the perturbation by the secondary.

\section{Coherence of periastra and eccentricities \label{sec:coherence}}

Relative shape and alignment of neighbouring orbits is a very
important factor because it controls the relative velocities of the
particles and thus their growth rate. 
Even if the companion star excites high eccentricities, it does not
necessarily mean that the relative velocities are high.
\citet{Marzari00} have shown that secular perturbations from an
eccentric companion, combined with the gas drag, lead to the strong
alignment of periastra between particles of the same size on
neighbouring orbits.
Since periastra are also coupled to eccentricities, the relative
velocities are low. This effect of ``orbital phasing'' in an
eccentric binary is prominent even for bodies of 100 km in diameter,
which are commonly regarded as decoupled from the gas. 
It should be stressed that the collision velocities are low
only between particles of the same size. \citet{Thebault06} has shown
that for distribution of sizes, even small misalignment in the periastra of
particles of different sizes results in relative velocities that are high enough
to prohibit collisional growth (note however that they used an equilibrium
axisymmetric-disk approximation).

In this section we investigate the effects of the orbital
coherence, but in the circular binary system and for smaller sizes of
particles. The secular effects of eccentricity forcing and periastra
alignment are not present in the circular system.
However, the inclusion of a non-axisymmetric gaseous 
wave pattern provides an additional factor that localy perturbs orbit
of the particle. Since the wave pattern co-rotates with the companion
star, we may expect similar effects to the secular gravitational
perturbations. To check what really happens in
such systems, we performed 4 runs with non-interacting particles of
sizes 1\,m, 10\,m, 100\,m, and 1\,km. Each run was initiated with 30000
particles on circular, non-inclined orbits distributed randomly between
0.8\,AU and 6\,AU.

\begin{figure}
\resizebox{\hsize}{!}{
\includegraphics[]{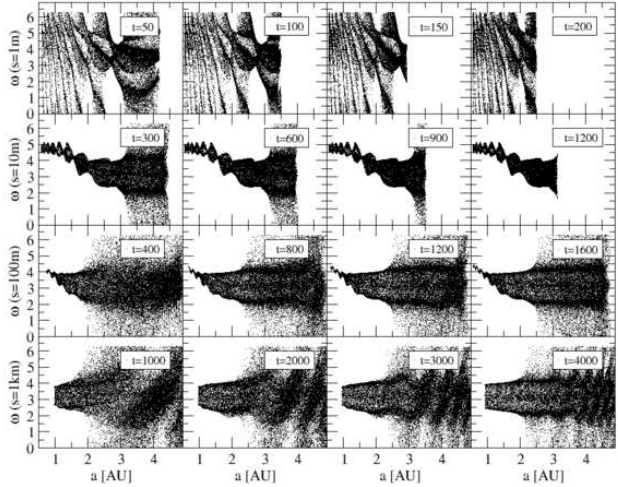}
}
\caption{\label{fig:w-a} Periastron longitude $\tilde{\omega}$ in
model W2. Each row shows temporal evolution for a particle of
a given size. The time is displayed in years.}
\end{figure}

Figure \ref{fig:w-a} shows the longitudes of particle periastra with
respect to the longitude of the companion, $\tilde{\omega}$, as a
function of particle semimajor axis. 
Each row shows the evolutionary sequence for the population of particles 
of different size (indicated on the vertical axis legend).
The time evolution is shown to prove that the observed
configurations are not transient but converge to a certain,
stationary pattern on the $\omega-a$ plane. 

A quick look at the plot reveals that the effect of periastra alignment
{\em is}  present for all considered sizes of particles.
The degree of alignment depends, however, on the distance from the
central star and on the particle size.
A closer exploration of the plot allows two different regimes to be
distinguished depending on the particle size:
\begin{enumerate}
\item Particles with radii smaller than a few meters feel strong
gas drag, and their periastra are correlated with the spiral pattern in
the gaseous disk. It can be observed on the plot for 1\,m particles in
the region of stable orbits (below ~3\,AU). 
\item Particles with radii larger than a few meters have orbits
aligned, but the alignment is not correlated with the waves in the
gaseous disk. In the outer parts of the disk, around 3\,AU, the periastra
are in opposition to the companion star with a substantial scatter in
longitude (larger for larger particles). When the drag force is
stronger (in the inner disk or for smaller particles),
the alignment is more pronounced. It is particularly strong for
10\,m particles.
\end{enumerate}

We want to stress here that the alignment of periastra does not
imply any spatial alignment of particles. In fact we did not observe a
spatial correlation of particles with the spiral waves at the particle-size
range considered here. We only checked that such a correlation becomes 
weakly visible for 10\,cm particles.

\begin{figure}
\resizebox{\hsize}{!}{
\includegraphics[]{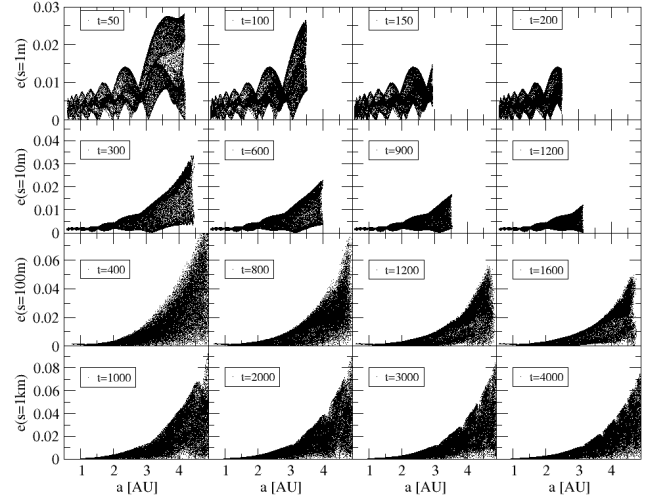}
}
\caption{\label{fig:e-a} Eccentricity as a function of semimajor axis.
Each row shows the temporal evolution for a particle of a given size.}
\end{figure}

Figure \ref{fig:e-a} illustrates the dependence of eccentricity on
semimajor axis in the same manner as for periastra longitudes. 
The coherence in eccentricity is much weaker in comparison to the
eccentric model of \citet{Marzari00}. In the inner disk, it is simply
the effect of strong damping by the gas drag.
The presence of spiral waves manifests mainly for 1\,m and
10\,m size particles in the form of pulsations in $a$. Actually, two
pulsation patters can be noticed for 1\,m particles corresponding to
two spiral arms. We expect that the relative phase between those
two patterns depends on the winding angle of the spirals. 
Thus the coherence in eccentricity will vary depending on the disk
temperature. 

\section{Discussion and summary \label{sec:summary}}

We have investigated orbital evolution of particles moving in a
circumprimary gaseous disk in a circular binary system. 
This is the first investigation of the orbital motions that includes
tidally excited density waves in the gas-drag calculation. 

We have demonstrated numerically that the radial flow of gas in the
disk (inward or outward) effectively increases the drag force due to
an enhanced mass flux colliding with the solid particles.
We derived approximate analytical formulae for the change rate of
orbital elements ($a, e, i$) in the gaseous disk, which is not in
radial equilibrium. The formulae do not assume anything about the
source of the radial gas flow. In general there are four constituents of
the relative velocity between particle and gas. The first
two, eccentricity and inclination, describe the deviation of the particle
from the Keplerian gas flow. The remaining two represent the radial
and angular deviation of the gas from the Keplerian flow. To account
for the non-axisymmetric features in the disk, one has to introduce {\em
effective} components of the gas velocity. The 
effective components  can be substantially larger than simple angular
averages, meaning that the evolution of orbital elements can be faster in
the non-axisymmetric disk. Exactly this situation happens in the
investigated disk perturbed by the companion star: the tidally induced
spiral waves which propagate radially in the form of shocks increase
the effective components of the gas velocity. In particular, in the outer disk
where the perturbation is strongest, the effective radial component
dominates other components of the relative velocity. We have
found numerically that the particles of sizes from 1\,m to 10\,km
migrate there around three times faster than in an axisymmetric disk
in radial equilibrium. This result agrees with the analytical
prediction very well, within a few percent.

The effect of enhanced drag force due to radial gas flow has some
significant consequences in the context of planetesimal 
formation in the binary system. The faster particle migration raises
the old question of whether there is enough time to form planetesimals
before all smaller particles have fallen onto the star.
We have to note, however, that the enhancement in the drag force
gradually vanishes inward in the disk. Depending on the disk
temperature and density radial profile, this differential migration may
eventually result in the accumulation of particles at a certain radius in
the inner disk, although we estimate that this is not the case for
realistic disk models.
There are other effects that may support faster planetesimal
formation. If only the larger bodies can form quickly, the
size-independent enhancement of the migration speed will result in
a higher flux of smaller particles, which will feed the larger body.
Furthermore, an accelerated damping of the particle's inclinations
leads to an increase in their number density in the mid-plane of the disk
and more frequent collisions. Both effects are important for
the planetesimal formation and early stages of their growth.

The frequency of collisions between particles is only one of the two
factors controlling their growth rate. The other one is the relative
velocity of colliding bodies that determines whether the bodies stick
together or shatter into smaller pieces. The relative velocity almost
exclusively depends on the eccentricity and periastra orientation of
the particles, provided that the inclinations are already damped. 
The contribution from decay of the semimajor axis is negligible 
(in our system it can play a role only for 1\,m particles at 3\,AU). 
We find that the spiral waves induce a certain coherency in both
periastra longitudes and eccentricities. Only for the smallest particles
of 1\,m size does the coherence in periastra longitudes come from a
simple correlation of the orbit orientation with the spiral wave pattern in
the gas. For larger bodies, there is no such correlation, and the degree
of coherence decreases with increasing size of the body. 
It seems that the relative velocities will be affected by coherence
only for bodies smaller than 10\,m.
In order to determine how much it does help in planetesimal
formation, collisional simulations with realistic size distribution
will have to be carried. This will be the subject of the next paper.

In this paper we focused on the circular binary system because it
allows the sole effects of the spiral waves to be studied without 
time dependent effects that are present in the eccentric binary. On
the other hand, the radial gas flow in the eccentric system is much
stronger and we expect it to influence particle motion to a much
higher degree than in the circular case. The eccentric binary case
will be investigated in one of the subsequent papers.

\begin{acknowledgements}
P.C. acknowledges financial support provided through the European
Community's Human Potential Programme under contract
HPRN-CT-2002-00308, PLANETS. P.C. and A.G. were supported by the
Polish Ministry of Science through grant No. 1 P03D 026 26.
The software used in this work was in part developed by the
DOE-supported ASC / Alliance Center for Astrophysical Thermonuclear
Flashes at the University of Chicago.
\end{acknowledgements}

\renewcommand{\theequation}{A.\arabic{equation}}
\setcounter{equation}{0} 

\begin{appendix}
\section{Appendix: Mean variations of orbital elements}
In this section we extend calculations of A76 for the case of a disk in
radial non-equilibrium. All other assumptions made in A76 are
preserved; in particular, the disk is axisymmetric. 
Parts of the formulae written in bold font are additions with respect
to their formulae.

Introducing non-dimensional radial velocity $\kappa=-V_r/\vk$,
Eqs. (4.7) in A76 take the following form:
\begin{eqnarray}
\label{eq:dadt-ext}
\frac{\rd a}{\rd t} &=& -A\rho u \frac{2a}{1-e^2} \left\{ 1+2e \cos\psi + e^2
- (1+e \cos\psi)^{3/2}h \cos i \right. + {}\nonumber\\ 
& & {} + \left. \boldsymbol{\kappa (1-e^2)^{1/2}e\sin \psi} 
\right\}, \\
\label{eq:dedt-ext}
\frac{\rd e}{\rd t} &=& -A\rho u \left\{ 2\cos \psi+2e-
\frac{2\cos\psi + e + e \cos^2\psi}{(1+e\cos \psi)^{1/2}}h\cos i
+ \right. {}\nonumber\\ 
& & {} + \left. \boldsymbol{\kappa (1-e^2)^{1/2}\sin \psi}
\right\}, \\
\label{eq:didt-ext}
\frac{\rd i}{\rd t} &=& -A\rho u \frac{h}{(1+e\cos\psi)^{1/2}}
\cos^2(\psi+\omega)\sin i ,
\end{eqnarray}
where $\psi$ and $\omega$ denote true anomaly and argument of
periastron respectively, and $h \approx 1-\eta$.
The total relative velocity $u$ reads:
\begin{eqnarray}
\label{eq:usq}
u^2 &=&\vk^2(a)\{(1-\frac{3}{4}\cos^2
\psi)e^2+\cos^2(\psi+\omega)i^2+\eta^2+\eta e\cos \psi\} + {}\nonumber \\
& & {} \boldsymbol{+ \frac{\vk^2}{1-e^2}\left\{ 
\kappa^2(1-e^2)-2\kappa e(1-e^2)^{1/2}
\sin \psi \right\}}.
\end{eqnarray}
Assuming that the drag force is weak and the orbital elements are
constant during one orbital period, their orbitally averaged rate of
variation is expressed as
\begin{equation}
\label{eq:orbavg}
\left\langle\frac{\rd Q}{\rd t}\right\rangle = \frac{1}{2\pi}\int_{0}^{2\pi}
\frac{\rd Q}{\rd t}\frac{(1-e^2)^{3/2}}{(1+e\cos \psi)^2} \mathrm{d}\psi,
\end{equation}
where $Q\in\{a,e,i\}$.
In order to simplify further calculations we apply the approximation
(Eq. 4.19 in A76):
\begin{equation}
\label{eq:uF_approx}
\langle uF \rangle = \langle u \rangle \langle F \rangle =
\left\{\langle u^2 \rangle\right\}^{1/2} \langle F \rangle
\end{equation}
where $F$ denotes factors other than $u$ in Eqs.
(\ref{eq:dadt-ext})-(\ref{eq:didt-ext}).
It is justified as long as $u$ and $F$ are independent and the constant
part of $u^2$ is greater than the amplitude of the oscillatory part.
Providing that variables $e,i,\eta,\kappa$ are much smaller than unity 
and preserving only leading terms for each of them, we obtain from 
Eq. (\ref{eq:usq})
\begin{equation}
\langle u^2 \rangle = \vk^2(a)\left[ 
\left(\frac{5}{8}-\boldsymbol{\kappa^2}\right) e^2 + \frac{1}{2}i^2+
\eta^2+ \boldsymbol{\kappa^2}
\right].
\end{equation}
In practice, term $\kappa^2$ in front of $e^2$ is negligible and the
remaining formula is intuitive: in addition to the angular component of
velocity, $\eta$, its radial part, $\kappa$, enters.

Fortunately, the orbital average of $F$ is not changed with respect
to A76, since all new terms in
Eqs. (\ref{eq:dadt-ext})-(\ref{eq:didt-ext}) vanish when
averaged. Thus the only changes come from $<u^2>$ term and finally  
we obtain:
\begin{equation}
\label{eq:dadt}
\frac{\tau_0}{a}\left\langle\frac{\rd a}{\rd t}\right\rangle = 
-2\left[ \left(\frac{5}{8}-\kappa^2\right) e^2 + 
\frac{1}{2}i^2+ \eta^2+ \kappa^2 \right] ^{1/2}\eta
\end{equation}
\begin{equation}
\label{eq:deidt}
\frac{\tau_0}{e}\left\langle\frac{\rd e}{\rd t}\right\rangle = 
2\frac{\tau_0}{i}\left\langle\frac{\rd i}{\rd t}\right\rangle = 
\left[ \left(\frac{5}{8}-\kappa^2\right) e^2 + 
\frac{1}{2}i^2+ \eta^2+ \kappa^2 \right] ^{1/2}.
\end{equation}

\end{appendix}

\end{document}